\documentclass[aps,prb,preprint,groupedaddress]{revtex4}

\usepackage{graphicx}
\usepackage{verbatim}
\usepackage{amsmath}
\usepackage{amsfonts}
\usepackage{amssymb}
\usepackage[usenames,dvipsnames]{color}

\begin{document}


\title{Enhancement of quantum dot luminescence in all-dielectric metamaterial}


\author{Vyacheslav V. Khardikov}
\email[]{khav77@gmail.com}
\affiliation{Institute of Radioastronomy, National Academy of
Sciences of Ukraine, 4, Krasnoznamennaya Street, Kharkiv 61002,
Ukraine} \affiliation{School of Radio Physics, Karazin Kharkiv
National University, 4, Svobody Square, Kharkiv 61077, Ukraine}

\author{Sergey L. Prosvirnin}
\email[]{prosvirn@rian.kharkov.ua}
\homepage[]{http://ri.kharkov.ua/prosvirn/}
\affiliation{Institute of Radioastronomy, National Academy of
Sciences of Ukraine, 4, Krasnoznamennaya Street, Kharkiv 61002,
Ukraine} \affiliation{School of Radio Physics, Karazin Kharkiv
National University, 4, Svobody Square, Kharkiv 61077, Ukraine}



\begin{abstract}
We propose a simple design of all-dielectric silicon-based planar
metamaterial manifested an extremely sharp resonant reflection and
transmission in the wavelength of about 1550~nm due to both low
dissipative losses and involving a trapped mode operating method.
The quality factor of the resonance exceeds in tens times the
quality factor of resonances in known plasmonic structures. The
designed metamaterial is envisioned for aggregating with a pumped
active medium to achieve an enhancement of luminescence and to
produce an all-dielectric analog of a "lasing spaser". We report
that an essential enhancement (more than 500 times) of luminescence
of layer contained pumped quantum dots may be achieved by using the
designed metamaterial. This value exceeds manyfold the known
luminescence enhancement by plasmonic planar metamaterials.
\end{abstract}




\maketitle


\newcommand{\sergei}[1]{ \textcolor{Turquoise}{({\bf Comment from Sergei:} #1)}}

\section{Introduction}
\label{sec:introduction}

Modern nanotechnologies effort an opportunity to structure optically
thin layers of materials with a periodic subwavelength pattern in
order to produce planar metamaterials. The planar metamaterials are
an impressive contemporary object, which is driven by certain
fascinating facilities both mostly well known over the decade and
quite novel. Recent works report results of aggregating laser
materials with planar metamaterials to design parametric gain
systems \cite{wegener-2008-tmf,fang-2009-scc,shalaev-2009-fds} and
to develop gaining or lasing devices such as the spaser
\cite{bergman-2003-spaser,zheludev-2008-lspaser,ziolkowski-2012-ios}.
An essential part of this development shell be the study of
luminescence of an active material hybridized with a planar
metamaterial that could support a coherent high-Q electromagnetic
oscillation.

Typically, a planar metamaterial assigned for visible and near
infrared wavelengths is a plasmonic structure designed on the basis
of a periodic array of complex-shaped resonant metallic nanowires or
slits in a metal slab. The main factor responsible for the
spectacular properties of these metafilms is some resonant
interaction of light with the patterned layer. Moreover, numerous
envisioned applications of planar metamaterials which incorporating
a pumped laser medium do require the high Q-factor resonance and a
strong confinement of electromagnetic fields.

However, losses of plasmonic metamaterials are orders of magnitude
too large for the envisioned applications. Typically, the Q-factor
of the resonances excited in plasmonic structures is small because
of high radiation losses and huge energy dissipation which is
inherent to metals in the visible and near infrared wavelength
ranges.

Fortunately there is a way of significantly decreasing of a
radiation losses level in the planar plasmonic structures. This way
lies in using so-called "trapped mode" resonances observed in planar
structures with broken symmetry of periodic elements due to exciting
of anti-phased currents in metallic elements of a subwavelength
periodic cell in microwaves \cite{prosvirnin-2003-rcm,
fedotov-2007-prl} and in an optical wavelength region
\cite{zhang-2008-prl, dong-2010-optexp, khardikov-2010-jop}.

Recently it was experimentally demonstrated that a hybridization of
semiconductor quantum dots (QDs) with a plasmonic metamaterial
supported the trapped mode resonance leads to a multifold intensity
increase and narrowing of their photoluminescence spectrum
\cite{tanaka-2010-meo}. The experiment shows that plasmonic
metamaterials interact with QDs like a cavity and luminescence
enhancement can be explained on base of the cavity quantum
electrodynamics Purcell effect. This observation is an essential
step towards understanding the mechanism of the radiation emission
in plasmonic metamaterials and opens amazing possibilities for
developing metamaterial-enhanced gain artificial media.

The trapped mode resonances have a typical peak-and-trough Fano
spectral profile \cite{fano-1961-pr} and can be excited, for
example, in the planar periodic array composed of twice
asymmetrically-split metallic rings. Their specific character arises
from destructive interference of the radiation by the anti-phased
current distribution. Theoretically, Q-factor of such resonance can
be infinitely increased by asymmetry degree decreasing in a
hypothetic lossless structure because the radiation loss level tends
to zero in this case. In practice, however, in visible and near
infrared, Q-factor strongly limited by energy dissipation which is
inherent to metals. Actually, the dissipative losses increase
extremely with increasing of currents in metallic elements with
decrease of radiation level from nearly symmetry elements of an
array. Thus the trapped mode resonances have a larger than a regular
resonance in the symmetry structure but still moderate Q-factor
\cite{khardikov-2010-jop}.

But there is a promising possibility to achieve a high-Q factor
trapped mode resonance in planar all-dielectric arrays with broken
symmetry of practically non-dissipative elements. It was shown in
\cite{khardikov-2012-jop} that using a planar array with square
periodic cell consisted of two different lengths semiconductor bars
enables to achieve the trapped mode resonance with Q-factor much
larger than related to this kind of resonances in plasmonic
structures. The Q-factor of the trapped mode resonance of the
germanium array can reach a value of the order of thousand in a near
infrared transparency window.

In the case of this array, the dielectric bars work as open
dielectric resonators and a pair of coupled resonators of each
periodic cell may be considered as a metamolecule. The trapped mode
resonance of a planar array is formed by destructive interference of
fields scattered by such metamolecules outside of the array. Besides
the large Q-factor, a remarkable property of this resonance of
all-dielectric arrays is an essential red shift of their resonant
frequency in comparison with a resonant frequency of an array which
periodic cell consists of a single dielectric bar (see
\cite{khardikov-2012-jop}). This property opens the way for using
dielectric materials with a relatively small refractive index to
produce this kind of arrays.

We can expect that a spectrum narrowing and intensity increasing of
the QDs photoluminescence observed in \cite{tanaka-2010-meo} will
rise with increasing the Q-factor of the planar structure resonance.
Thus it is extremely interesting to study features of
photoluminescence in a hybridization of QDs with a periodic
all-dielectric array in the trapped mode resonance regime.

\section{The problem statement and the method of solution}
\label{sec:theory}

\begin{figure}[htb]
\centerline{\includegraphics[width=10.0cm]{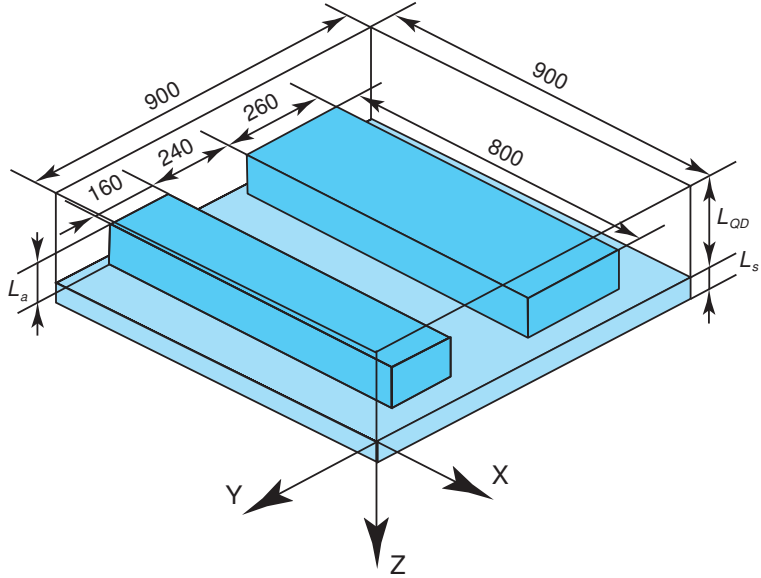}}
\caption{(Color online) A sketch of the unite cell of the
double-periodic planar structure. The all-dielectric array composed
of two dielectric bars per a periodic cell is immersed into the QD
layer.} \label{fig:fig1}
\end{figure}

Let us consider a double periodic array composed of dielectric bars
placed on a substrate with thickness $L_s$ (see
Fig.~\ref{fig:fig1}). The periodic array of thickness $L_a$ is
immersed into QD layer (polymer layer in which QDs are dispersed)
with total thickness $L_{QD}$. The unit cell of the array includes a
pair of dielectric bars which have different sizes. In the paper
\cite{khardikov-2012-jop} asymmetry of double periodic array of
dielectric elements was obtained by using dielectric bars of
identical square cross sections but different lengths. But it is
more suitable for practice using the dielectric bars differed in
width sizes along $y$-direction. If all bars will be identical in
both the length and the thickness then such double periodic array is
very convenient for making. The sizes of the square periodic cell
are chosen $d_x=d_y=d=900$~nm. The periodic cell is assumed to be
symmetric relatively to the line drawn through the cell center and
parallel to the $y$-axis. The normal incidence of a linearly
$x$-polarized plane wave is considered.

As material for the dielectric bars we propose to use the silicon as
most widespread material of electronics which has a transparency
window in the wavelength range from 1150~nm to 2500~nm. The sizes of
the array elements are chosen to provide the trapped mode resonance
of optically non-pumped structure in wavelength 1550~nm, which is
the typical wavelength used in telecommunications. In the wavelength
1550~nm, a silicon refractive index is approximately 3.48. Using an
extremely thin substrate enables to exclude undesired wave
interference. The substrate is assumed to be a synthetic fused
silica membrane. Its refractive index is approximately $n_s=1.45$ in
the wavelength range under consideration \cite{malitson-1965-josa}.
The substrate thickness is 50~nm. It needs to notice that there are
not high non-evanescent diffraction orders in the substrate (an
inequality $n_s d <1550$~nm is valid).

To study an intensity enhancement of the luminescence of QDs
resulted from their hybridization with a dielectric array we propose
to use the solutions of plane wave diffraction problem in the cases
of the structure with and without optical pump. In proposed method
the photoluminescence emission intensity $W_{e}$ is evaluated as a
difference between intensity of dissipation (or emission) defined by
expression $1-|R|^2-|T|^2$ in the structure without optical pump and
with it
\begin{equation}\label{eq:one}
W_e=W_{0}-W_{+},
\end{equation}
where $W_{0}$ and $W_{+}$ are intensities of dissipative losses (or
emission) in the structure without and with optical pump
respectively, $R$ and $T$ are coefficients of reflection and
transmission. It needs to notice that the value $W=1-|R|^2-|T|^2$
will be positive if a dissipation is observed in the structure and
negative in the case of a total intensity ($|R|^2+|T|^2$) of
reflected and transmitted waves exceeds the intensity of an incident
wave. This fact was taken into account in formula (\ref{eq:one})
where the photoluminescence intensity $W_{e}$ has a positive value.

The proposed method was tested for validity on an example of a
plasmonic metamaterial combined with QDs which was studied
experimentally in \cite{tanaka-2010-meo} (see Appendix). We use the
experimental data on a luminescence spectrum presented in
\cite{tanaka-2010-meo} to define parameters for a theoretical
adequate description of an actual composite material which is QDs
dispersed in a host polymer layer. In additional, the comparison of
experimental results of this work and our theoretical ones give us
an evidence of validity of the proposed theoretical approach to
study photoluminescence.

Thus to determine the photoluminescence intensity it is required to
solve the plane wave diffraction problem for considered structures.
For this aim the mapped pseudospectral time domain method proposed
in \cite{gao-2004-osa, khardikov-2008-radio} is used. For the sake
of simplicity the dispersion of dielectrics is not taken into
account in this paper. There are two reasons for using such
approximation. First, the dispersion of the chosen materials is very
weak in the considered wavelength range. Next this, an actual
dispersion has no effect on the properties of the trapped-mode
resonances.

To take into account a gain in optically pumped QDs the method of
additional differential equations \cite{taflove-2005-fdtd} was used
with a model of negative frequency-dependent conductivity
\cite{hagness-1996-radsci}. The expression of corresponding
time-domain conductivity $\sigma(t)$ is chosen in the form to ensure
that this value is real and causal
\begin{equation}\label{eq:two}
 \sigma(t)=\frac{\sigma_0}{\tau}\cos({\omega_0t})e^{-t/\tau}u(t).
\end{equation}
Here $u(t)$ is the Heaviside unit step function. The coefficient
$\sigma_0$ is proportional to the peak value of the gain set by the
pumping level and the resulting population inversion. Such model
assumes that the optical gain medium is homogeneously broadened
wherein the atoms of the gain medium are indistinguishable and have
the same atomic transition frequency, $\omega_0$. A natural
variation of the QD sizes will cause the broadening of the exciton
line of the QDs. Time constant $\tau$ permits to include the
relaxation processes in a phenomenological manner (decay rate is
$1/\tau$) and shows that any phase coherence introduced into the
system of atoms by the action of an electric field will be lost in a
time interval $\tau$ once the electric field is turned off.

The Fourier transform of $\sigma(t)$  reduces to following
frequency-dependent conductivity
\begin{equation}\label{eq:three}
 { \sigma(\omega) = \frac{\sigma_0/2}{1+i(\omega-\omega_0)\tau}
 +\frac{\sigma_0/2}{1+i(\omega+\omega_0)\tau}
   =
   \frac{\sigma_0(1+i\omega\tau)}{(1+\omega_0^2\tau^2)+2i\omega\tau-\omega^2\tau^2},}
\end{equation}
where a time dependence of electromagnetic field is chosen in the
form $\exp(i\omega t)$. It is clear from (\ref{eq:three}) that the
gain coefficient is governed by a single Lorentzian profile having a
width determined by $\tau$. The resonant frequency, that is the
frequency at which the response is maximized, is given by
$\omega_d=\sqrt{\omega_0^2+\tau^{-2}}$. The peak of the gain curve
is $\sigma(\omega_d)=\sigma_0/2$ and the full-width-at-half-maximum
bandwidth is $\delta\omega=2/\tau$. Using (\ref{eq:three}) one can
obtain a complex propagation constant of a plane wave and see that
the wave amplification will be observed only for the case of
$\sigma_0<0$.

The chosen form of frequency-dependent conductivity may be easily
included in the calculation scheme of time domain approaches by
using of ADE method. It results in two additional first order
differential equations in time in each grid node immersed into the
gain medium \cite{hagness-1996-radsci}.

A good agreement of our theoretical and known experimental results
of study luminescence of plasmonic metamaterial combined with QDs
(see Appendix) is an evidence of correctness both the model of
active QD medium and the proposed approach to study luminescence in
a planar resonant metamaterial.

Now let us apply the proposed method to design a silicon resonant
metamaterial combined with semiconductor QDs to enhance the
photoluminescence intensity with the spectral maximum in wavelength
1550~nm.

For numerical study, a laser medium based on semiconductor QDs was
chosen with following parameters:
$\omega_0=1.26\cdot10^{15}$~$\textrm{s}^{-1}$ which corresponds to
wavelength $\lambda_0=1550$~nm; $\tau=4.85\cdot10^{-15}$~s;
$\epsilon_{QD}=2.19$ which corresponds to refractive index
$n_{QD}=1.48$ of non-pumped QD laser medium, and
$\sigma_0=-500$~Sm/m corresponding to an emission factor
$\tan\delta_e=-0.021$ on the analogy of a lossy factor of media.
Small value of $\tau$ results in a wide-band QD spectral line and it
enables to exclude from consideration the effects caused by
displacement of metamaterial dissipation peak and maximum of exciton
emission line of QDs. Let us notice that the pump level (parameter
$\sigma_0$) is in one order less than it was need in the case of
plasmonic metamaterials (see Appendix) because of low losses of
all-dielectric array.

The geometry parameters of the dielectric bars array immersed in
non-pumped QD layer ($\sigma_0=0$) were chosen to provide a high
Q-factor trapped mode resonance of the structure near the wavelength
1550~nm. Parameters of the silicon structure corresponding to this
condition are following $L_a=100$~nm, $L_{QD}=210$~nm, $L_s=50$~nm,
length of the dielectric bars is 800~nm, widths of bigger and
smaller bars are 260~nm and 160~nm respectively, and the distance is
240~nm between two different width bars (see Fig.~\ref{fig:fig1}).

To take into account the energy dissipation in dielectric bars, a
model of constant conductive medium was used for a description of
dielectric in the PSTD method. It means that the dielectric of bars
is modeled as medium with both $\epsilon_{a}=n_{a}^2$ and
$\sigma_{a}$ being some constants. Such approach results in
frequency-dependent losses.

\section{Analysis of luminescence of hybridization of QDs with
a low-loss all-dielectric planar metamaterial} \label{sec:res1}

\begin{figure}[htb]
\centerline{\includegraphics[width=10.0cm]{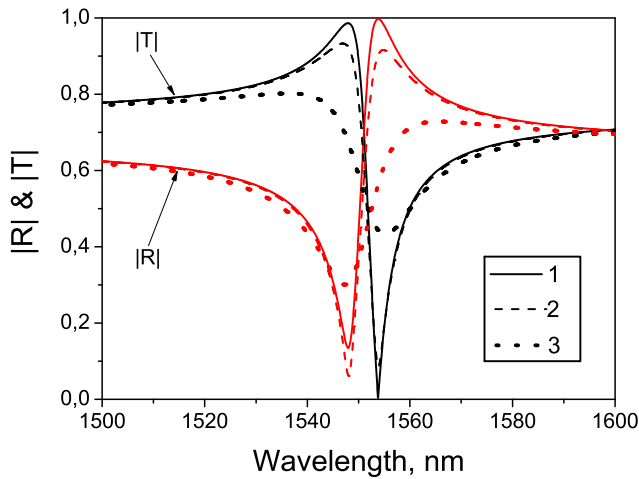}}
\caption{(Color online) Wavelength dependences of reflection and
transmission coefficients of the periodic array of silicon bars
immersed in the non-pumped QD layer. Lines~1, 2, and 3 correspond to
$\tan\delta_{a}=0,$ $10^{-3}$, and $10^{-2}$ respectively.}
\label{fig:fig2}
\end{figure}

The wavelength dependences of reflection and transmission
coefficients magnitudes of the designed structure are shown in
Fig.~\ref{fig:fig2}. There is a trapped mode resonance near the
wavelength 1550~nm which has the typical trough-and-peak Fano
spectral profile. As it was shown in \cite{khardikov-2012-jop}, this
resonance results from the anti-phased excitation of the dielectric
bars operated as a half-wavelength dielectric open resonators in
this case. Lines~1, 2, and 3 correspond to $\sigma_{a}=0$,
$\sigma_{a}=130.3$~Sim/m, and $\sigma_{a}=1303$~Sim/m or
equivalently $\tan\delta_{a}=0$, $\tan\delta_{a}=10^{-3}$, and
$\tan\delta_{a}=10^{-2}$ in the wavelength 1550~nm. Let us notice
that in reality the silicon lossy factor $\tan\delta_{Si}$ is less
than $10^{-3}$ in this wavelength.

We estimate the trapped mode resonance Q-factor by using the
following formula proposed in \cite{khardikov-2012-jop}
\begin{equation}\label{eq:four}
 Q=\frac{\lambda_1\lambda_2}{\lambda_0|\lambda_2-\lambda_1|}=
 \frac{2\lambda_1\lambda_2}{|\lambda_2^2-\lambda_1^2|},
\end{equation}
where $\lambda_1$ and $\lambda_2$ are wavelengths of the maximum and
minimum of reflection or transmission coefficient of corresponded
Fano spectral profile and $\lambda_0=(\lambda_1+\lambda_2)/2$ is an
average wavelength of the trapped mode resonance. Using
(\ref{eq:four}) it was obtained following Q-factor values 268, 232
and 83 corresponded to $\tan\delta_{a}=0$, $10^{-3}$ and $10^{-2}$.
It is interesting that wavelength of maximum and minimum of
reflection and transmission coefficients coincide only in the case
of non-dissipative structure. Reflection and transmission extremes
are shifted relatively to each other when dissipation in dielectric
taken into account and then the Q-factor of trapped mode resonances
is estimated as the average value $Q=(Q_R+Q_T)/2$. Here $Q_R$ and
$Q_T$ are calculated from wavelength dependences of reflection and
transmission coefficients by formula (\ref{eq:four}). Let us notice
that the average trapped mode resonant wavelength
$\lambda_0=((\lambda_0)_R+(\lambda_0)_T)/2=1551$~nm is observed in
all considered cases of dissipative losses.

\begin{figure}[htb]
\centerline{\includegraphics[width=10.0cm]{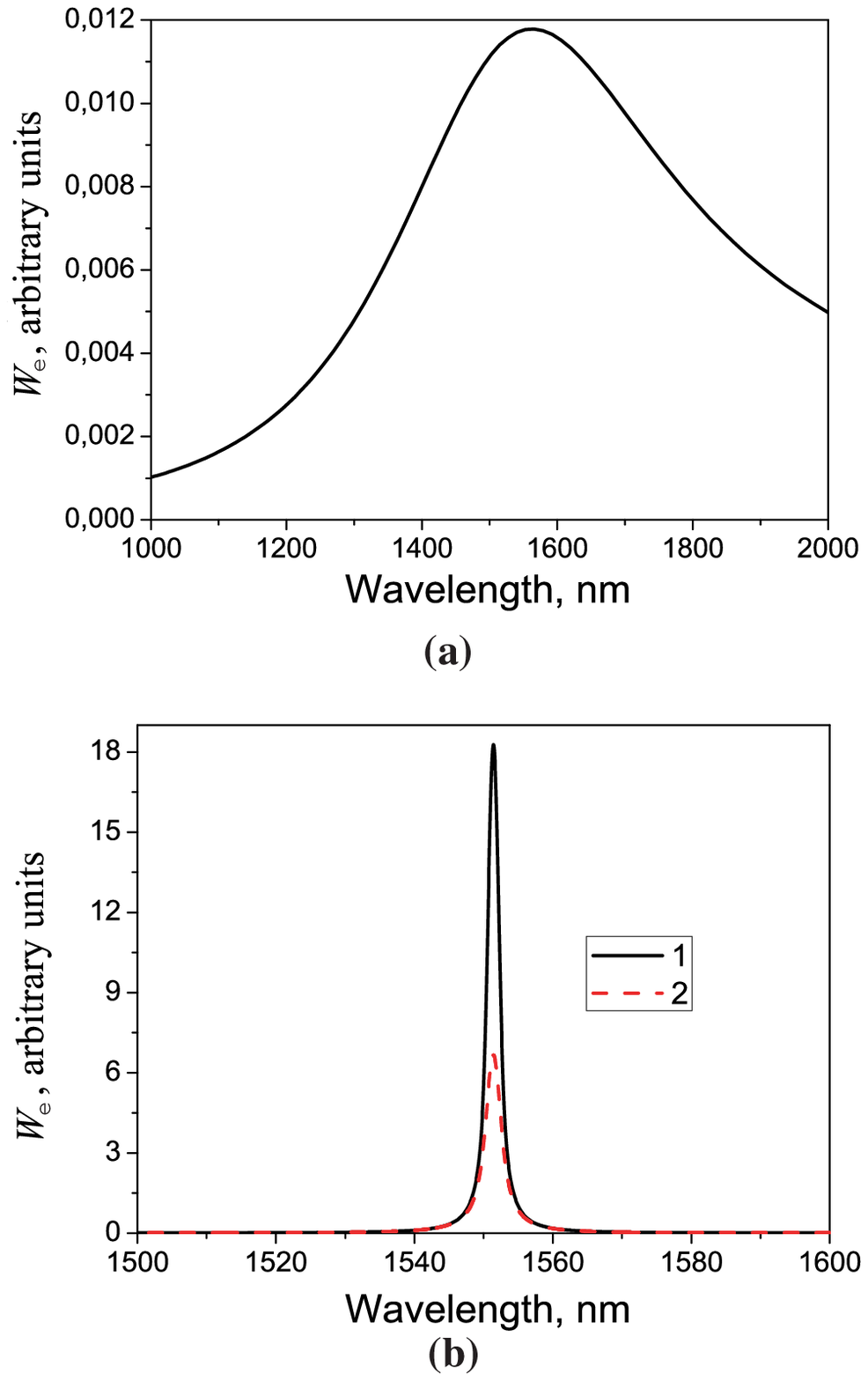}}
\caption{(Color online) Wavelength dependences of luminescence
intensity of QD layer with thickness 210~nm placed on silica
membrane (a), and QD layer aggregated with metamaterials (b): 1 -
$\tan\delta_{Si}=10^{-3}$, 2 - $\tan\delta_{a}=10^{-2}$.}
\label{fig:fig3}
\end{figure}

The wavelength dependences of photoluminescence intensity $W_{e}$
calculated by using (\ref{eq:one}) are presented in
Fig.~\ref{fig:fig3}. Fig.~\ref{fig:fig3},a shows the
photoluminescence intensity of homogeneous QD layer which thickness
is $L_{QD}=210$~nm placed on 50~nm thick silica membrane i.e.
intensity related to layered structure without any array. The
wavelength dependencies of photoluminescence of QD hybridization
with non-dissipative and low-dissipative ($\tan\delta_{a}=10^{-3}$)
metamaterials are presented in Fig.~\ref{fig:fig3},b. One can see a
large enhancement of photoluminescence intensity. For the case of
non-dissipative bars the photoluminescence intensity enhances in
1560 times. If the dissipation of silicon estimated as
$\tan\delta_{Si}=10^{-3}$ is taken into account, this coefficient of
intensity enhancement will be 560. These values are much bigger than
known orders of photoluminescence enhancement inherent to plasmonic
metamaterials combined with QDs (see \cite{tanaka-2010-meo} and
Appendix).

\begin{figure}[htb]
\centerline{\includegraphics[width=10.0cm]{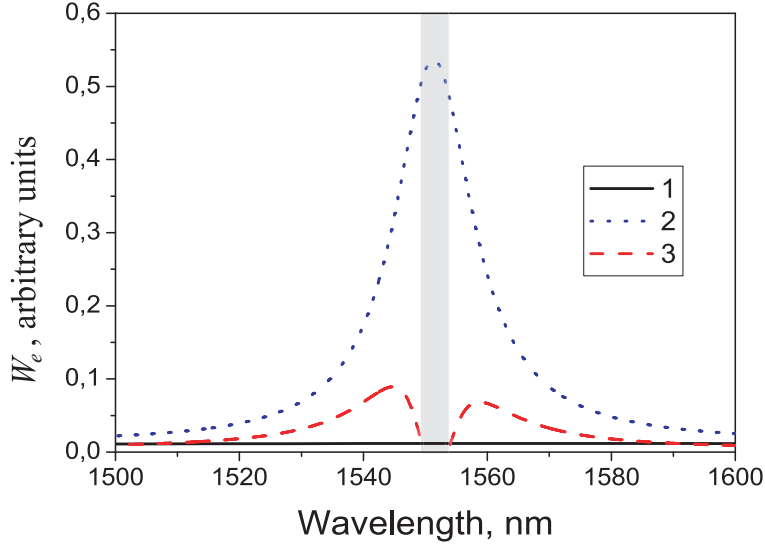}}
\caption{(Color online) Wavelength dependences of photoluminescence
of QD layer with thickness 210 nm (line 1), energy absorption of QD
without optical pump aggregated with array of silicon bars (line 2)
and photoluminescence of metamaterials aggregated with QD (line 3).}
\label{fig:fig4}
\end{figure}

Up to this point we consider photoluminescence in the pumped
all-dielectric metamaterial with emission factor
($\tan\delta_e=-0.021$) which exceeds essentially in absolute value
the lossy factor of metamaterial ($\tan\delta_a=10^{-3}$). Now let
us study the case of pumped lasing medium with the intensity of
energy dissipation in the dielectric array ($\tan\delta_a=10^{-2}$)
is comparable with the gain of QDs. In Fig.~\ref{fig:fig4} we
present wavelength dependences of the photoluminescence intensity of
QD-layer placed on silica substrate (line~1), the power dissipation
in the array of dielectric bars hybridizing with non-pumped QDs
(line~2), and the photoluminescence intensity of optically pumped
QDs aggregated with the same array (line~3). In the last case the QD
photoluminescence enhancement is observed too (see
Fig.~\ref{fig:fig4}, line~3). However, in the trapped-mode resonance
wavelength, we observe a decrease of emission intensity until
complete forbidding photoluminescence (see a grey filled region in
Fig.\ref{fig:fig4}, where $W_{e}<0$). Such wavelength dependence of
photoluminescence intensity of QDs aggregated with the array is
explained by fact that the QD gain results in an increase of field
amplitude inside dielectric bars and related increase of energy
dissipation in the array. In the case of low energy dissipation (for
example the dissipation of silicon array with
$\tan\delta_{Si}=10^{-3}$) this losses result in a simple decrease
of the level of photoluminescence enhancement.

\section{Conclusions}
\label{sec:conclusions}

We have proposed a simple design of all-dielectric silicon-based
planar metamaterial manifested an extremely sharp resonant
reflection and transmission in the wavelength of about 1550~nm due
to both low dissipative losses and involving trapped mode operating
method. Accurate numerical estimations make evidence that the
quality factor of the resonance of proposed structure exceeds in
tens times the quality factor of known plasmonic structures.

The designed planar metamaterial is envisioned for aggregating with
a layer of pumped active medium to achieve an enhancement of
luminescence and to produce an all-dielectric analog of the lasing
spaser \cite{zheludev-2008-lspaser}.

We have proposed an approach to study luminescence intensity of
pumped QDs aggregated with a planar metamaterial. The approach is
based on the comparison of results of two diffraction problems
solutions. They are the problems of plane electromagnetic wave
diffraction by the structure with and without optical pump. The
validity of the method has been argued by its application to
theoretical reproducing known experimental data on QD luminescence
in plasmonic metamaterials \cite{tanaka-2010-meo}.

We demonstrate that an essential enhancement (more than 500 times)
of luminescence intensity of layer contained pumped QDs may be
achieved by using the designed all-dielectric resonant metamaterial.
This value exceeds manyfold the known luminescence enhancement by
plasmonic planar metamaterials.

\section{Acknowledgments}
\label{sec:ack}

This work was supported by the Ukrainian State Foundation for Basic
Research, Project F40.2/037.

\section{\textbf{Appendix}\\Validation of diffraction approach to
study luminescence} \label{sec:res2}

\begin{figure}[htb]
\centerline{\includegraphics[width=10.0cm]{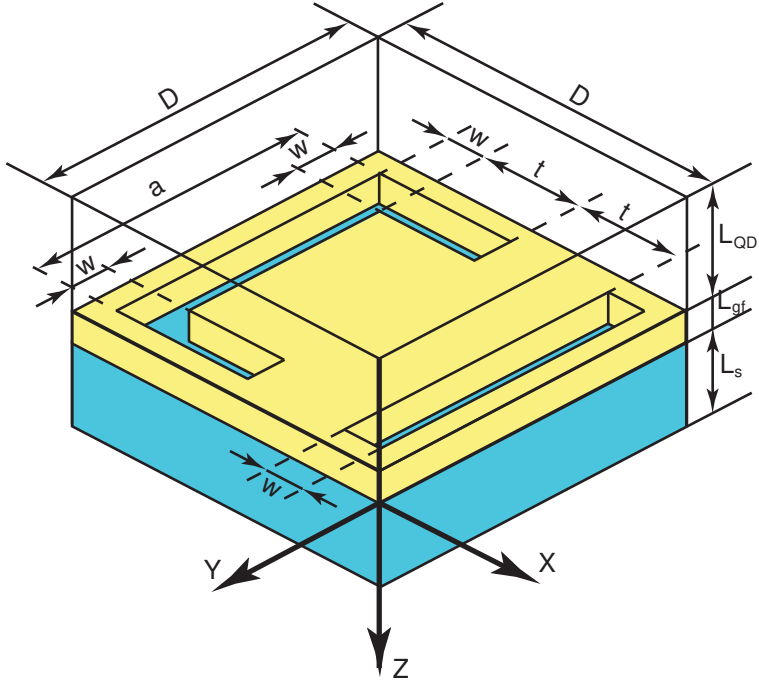}}
\caption{(Color online) A geometry of periodic cell of metamaterial
aggregated with QD/PMMA studied experimentally in
\cite{tanaka-2010-meo}.} \label{fig:fig5}
\end{figure}

For a validation of our approach to choosing parameters described
the medium being QDs dispersed in a host polymer and the diffraction
approach to study luminescence, let us considered one of the
plasmonic metamaterials which were treated experimentally in
\cite{tanaka-2010-meo}. The sample consists of a gold film patterned
periodically by asymmetric split-ring slits and placed on a glass
substrate (see Fig.~\ref{fig:fig5}). The film thickness is
$L_{gf}=50$~nm. A side size of a square periodic cell of the array
is $D=545$~nm. Sizes of the slits of a periodic cell shown in the
sketch in Fig.~\ref{fig:fig5} are following: $a=470$~nm, $t=170$~nm,
and $w=65$~nm. The slits are filled with a polymer material
contained semiconductor QDs and the array is coated by a layer of
the same material. The thickness of the layer is $L_{QD}=180$~nm. In
\cite{tanaka-2010-meo} was used lead sulfide (PbS) semiconductor QDs
from Evident Technologies with a luminescence peak around 1300~nm
and mean core diameter of 4.6~nm. These QDs were dispersed in
polymethylmethacrylate (PMMA).

Let us apply the diffraction approach based on the formula
(\ref{eq:one}) to estimate the photoluminescence intensity of
QD/PMMA hybridized with this plasmonic metamaterial.

The parameters of gain medium model must be determined before
luminescence intensity will be estimated. As it was mentioned above,
the following parameters $\omega_0$, $\tau$, and $\sigma_0$ are
required to determine a gain medium. Parameters $\omega_0$ and
$\tau$ of QD/PMMA may be determined from the photoluminescence peak
frequency $\omega_d=\sqrt{\omega_0^2+\tau^{-2}}$ and the
full-width-at-half-maximum bandwidth of QD/PMMA photoluminescence
$\delta\omega=2/\tau$. We use following values
$\omega_0=1.47\cdot10^{15}$~$\textrm{s}^{-1}$ and
$\tau=9.69\cdot10^{-15}$~s. Refractive index of non-pumped QD/PMMA
has the value of $n=1.48$ (see \cite{tanaka-2010-meo}). The value of
$\sigma_0$ depends on level of population inversion in QD/PMMA and
determines the intensity of photoluminescence. Here we use the value
$\sigma_0=-5000$~Sim/m to describe an optically pumped QD/PMMA. It
is a typical value which was used to describe a gain of
semiconductor media for example in \cite{hagness-1996-radsci}.

We consider the metamaterial aggregated with QD/PMMA placed on a
semi-infinite silica substrate ($L_s\rightarrow\infty$). Such
choosing enables us to exclude from the study the interference in
the substrate. The model of gold permittivity proposed in
\cite{vial-2008-appb} was used in our simulation of the plasmonic
metamaterial.

\begin{figure}[htb]
\centerline{\includegraphics[width=10.0cm]{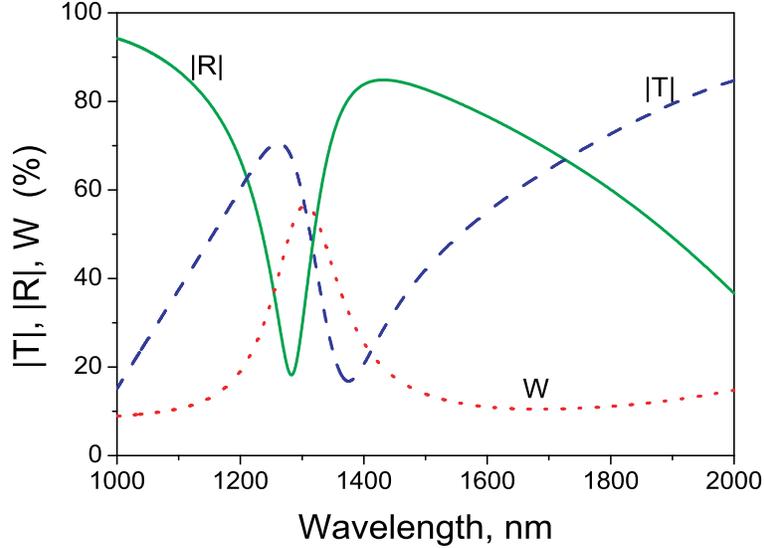}}
\caption{(Color online) Results of our numerical simulation  of
wavelength dependences of reflection, transmission, and dissipation
coefficients of the metamaterial aggregated with non-pumped QD/PMMA
researched experimentally in \cite{tanaka-2010-meo}.}
\label{fig:fig6}
\end{figure}

The results of our numerical simulation of wavelength dependences of
reflection, transmission, and dissipation coefficients of the
metamaterial aggregated with non-pumped QD/PMMA are shown in
Fig.~\ref{fig:fig6}. These theoretical results may be compared with
experimental ones presented in Fig.~1 of \cite{tanaka-2010-meo}. One
can see a very good agreement between our numerical and measured
results. Some difference of levels of reflection, transmission and
absorption observed in theoretical and experimental results can be
explained by distinctions of the actual structure from the perfect
periodic one.

\begin{figure}[htb]
\centerline{\includegraphics[width=10.0cm]{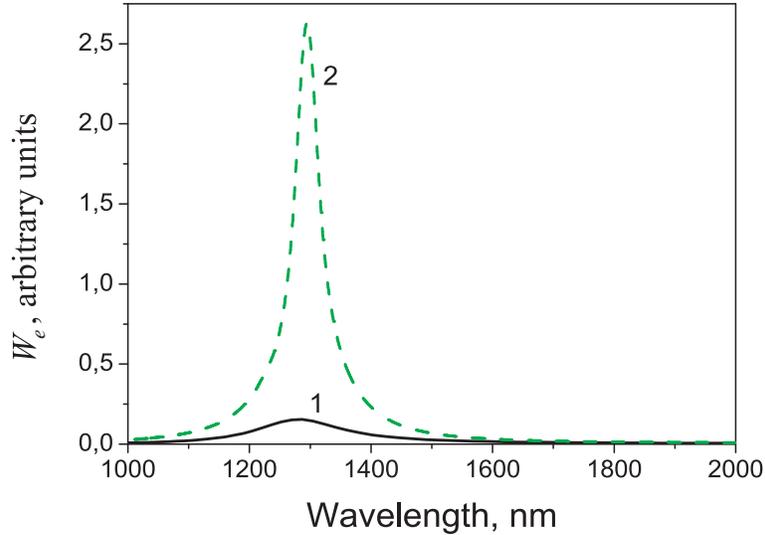}}
\caption{(Color online) Wavelength dependences of photoluminescence
intensity of homogeneous QD/PMMA layer (line~1) and the plasmonic
metamaterial aggregated with optically pumped QD/PMMA (line~2).}
\label{fig:fig7}
\end{figure}

To estimate an enhancement of luminescence intensity due to using
metamaterial structure, we compare its value with the luminescence
intensity of optically pumped QD/PMMA homogeneous layer placed on
the same substrate. The thickness of the layer is chosen the same as
a total thickness of the metallic array and its QD/PMMA coating. The
wavelength dependences of photoluminescence intensity of QD/PMMA
homogeneous layer which being thickness 230~nm (line~1) and
metamaterials aggregated with 180~nm-thick QD/PMMA layer (line~2)
are shown in Fig.~\ref{fig:fig7}. In this case we obtain 17 times
enhancement of photoluminescence intensity. This value of
luminescence enhancement exceeds only in twice the value obtained
experimentally in \cite{tanaka-2010-meo} that is quite good result
in modeling of active structures.

Thus, we have argued that the both our model of active QD medium and
the diffraction approach to study luminescence intensity result to
numerical data being in good correspondence with known measured
ones.

\bibliography{Luminescence-PRB}

\end{document}